\documentclass[10pt,journal]{IEEEtran}

\usepackage{balance} 

\usepackage{graphicx} 
\usepackage{float}
\usepackage{cite}
\usepackage{amsmath,amssymb}
\usepackage[outdir=./]{epstopdf}
\usepackage{afterpage}
\usepackage{optidef}
\usepackage{orcidlink}
\usepackage{soul}
\usepackage{tikz}
\usepackage{pgfplots}
\usepackage{pgfplotstable}
\usetikzlibrary{shapes.geometric, fit, arrows, positioning, backgrounds}
\usetikzlibrary{calc}
\usepackage{multirow}

\usepackage[font=footnotesize,caption=false]{subfig} 

\usepackage{comment}
\usepackage{optidef}

\usepackage{colortbl}  
\usepackage{tcolorbox} 
\usepackage{algorithm}
\usepackage{algpseudocode}
\usepackage{bm}
\usepackage{dsfont}

\DeclareMathOperator*\argmax{arg\,max}

\usepackage{xspace}

\usepackage[firstpage=true]{background}
\SetBgContents{\textcolor{black}{\footnotesize This work has been submitted to the IEEE for possible publication. Copyright may be transferred without notice, after which this version may no longer be accessible.}}
\SetBgPosition{current page.south}
\SetBgVshift{0.50cm}
\SetBgOpacity{1.0}
\SetBgAngle{0.0}
\SetBgScale{1.0}

\begin{document}
\bstctlcite{IEEEexample:BSTcontrol}

\title{Semantic-Aware Task Clustering for Federated Cooperative Multi-Task Semantic Communication  \thanks{This work was supported in part by the German Federal Ministry of Research, Technology, and Space (BMFTR) under Grant 16KISK016 (Open6GHub), and by the Deutsche Forschungsgemeinschaft (DFG, German Research Foundation) -- 518671822.}
}

\author{
\IEEEauthorblockN{Ahmad Halimi Razlighi\,\orcidlink{0009-0006-3826-832X}, Pallavi Dhingra\IEEEauthorrefmark{2}\,\orcidlink{0009-0005-5630-0488}, Edgar Beck\,\orcidlink{0000-0003-2213-9727}, Bho Matthiesen\,\orcidlink{0000-0002-4582-3938}, and Armin Dekorsy\,\orcidlink{0000-0002-5790-1470}} \\

\IEEEauthorblockA{Department of Communications Engineering, University of Bremen, Germany}\\

\IEEEauthorblockA{E-mails:\{halimi, beck, matthiesen, dekorsy\}@ant.uni-bremen.de}

\IEEEauthorblockA{\IEEEauthorrefmark{2}E-mail: pallavi1@uni-bremen.de
}

}

\maketitle

\begin{abstract}
Task-oriented semantic communication (SemCom) prioritizes task execution over accurate symbol reconstruction and is well-suited to emerging intelligent applications. Cooperative multi-task SemCom (CMT-SemCom) further improves task execution performance. However, \cite{halimi-letter} demonstrates that cooperative multi-tasking can be either constructive or destructive. Moreover, the existing CMT-SemCom framework is not directly applicable to distributed multi-user scenarios, such as non-terrestrial satellite networks, where each satellite employs an individual semantic encoder. In this paper, we extend our earlier CMT-SemCom framework to distributed settings by proposing a federated learning (FL) based CMT-SemCom that enables cooperative multi-tasking across distributed users. Moreover, to address performance degradation caused by negative information transfer among heterogeneous tasks, we propose a semantic-aware task clustering method integrated in the FL process to ensure constructive cooperation based on an information-theoretic approach. Unlike common clustering methods that rely on high-dimensional data or feature space similarity, our proposed approach operates in the low-dimensional semantic domain to identify meaningful task relationships. Simulation results based on a LEO satellite network setup demonstrate the effectiveness of our approach and performance gain over unclustered FL and individual single-task SemCom.
\end{abstract}


\section{Introduction} \label{section:Introduction}

Advances in artificial intelligence and end-to-end (E2E) learning-based system design have positioned \emph{semantic communication} (SemCom) as a key enabler for next-generation wireless networks, especially 6G, which are expected to support a wide range of intelligent applications \cite{taskcom6g}. Unlike conventional communication paradigms that aim at accurate symbol reconstruction, these emerging applications prioritize task completion. SemCom addresses this shift by transmitting only task-relevant information \cite{Gunduz2022}.

Within the task-oriented SemCom literature, existing works can be broadly categorized into single-task \cite{Shao2021,Beck2023}, and multi-task processing. Multi-task SemCom, improving model performance and generalization, further divides into non-cooperative and cooperative processing. Early studies consider non-cooperative settings where each task is learned independently using its respective datasets \cite{xie2022task}. More recent works apply established multi-task learning (MTL) architectures \cite{caruana1997multitask} to the SemCom domain. For instance, \cite{gong2023scalable} explores joint multi-tasking for SemCom systems exclusively based on machine learning (ML) approaches.

In contrast, previously we introduced an information-theoretic perspective on cooperative multi-tasking, moving beyond the black-box use of deep neural networks (DNNs) in \cite{halimi-letter}. We investigated a split structure for the semantic encoder, dividing it into a common unit (CU) and multiple specific units (SUs) to enable cooperative processing of multiple tasks. The resulting cooperative multi-task SemCom (\emph{CMT-SemCom}) framework facilitated multi-tasking based on a single observation. Moreover, subsequent works extended this framework to scenarios with distributed partial observations \cite{halimi_icc} and to rate-limited wireless channels \cite{halimi_ojcoms}.

However, directly applying the existing CMT-SemCom framework to distributed multi-user scenarios, such as non-terrestrial networks (NTNs), e.g., low-Earth orbit (LEO) satellite networks observing different regions of the Earth and operating with local datasets, is not feasible. This is due to the cooperation taking place at the transmitter side through the shared CU in the CMT-SemCom framework. LEO satellite networks are expected to play a central role in extending global connectivity and supporting intelligent services in future networks \cite{LEO5G}. Moreover, in these networks, a single satellite or constellation may simultaneously support multiple tasks, like agricultural monitoring, weather prediction, and emergency response. Recent advances in onboard computation and hardware capabilities have made it feasible to deploy SemCom directly on satellites, allowing communication resources to be allocated towards task-oriented objectives \cite{SemNTN}. Consequently, MTL and SemCom have recently become active research topics for satellite networks \cite{MT-Crosslayer,SemImgSatLEO,MT-SatImage}.  

Motivated by these insights, we propose federated learning (FL) to enable cooperative, multi-task semantic communication in distributed, multi-user networks, e.g., satellite networks. FL has been widely studied in satellite systems \cite{FLSat} as it is considered an important step towards seamless integration of terrestrial and non-terrestrial networks for supporting various applications. Recently, FL has also been combined with SemCom for single-task scenarios \cite{FLSemSat}. Further, recent works have begun investigating MTL via FL using purely ML-based approaches \cite{MT-FL,MT-FL2}. In this work, we extend our information-theoretic CMT-SemCom framework to distributed multi-user networks by incorporating FL as the mechanism for task cooperation. 

In addition, existing multi-task SemCom, including the CMT-SemCom, commonly assume that all tasks are fully related, and applying these methods to unrelated tasks leads to suboptimal task execution and potential performance degradation. As demonstrated in \cite{halimi-letter}, such a mismatched cooperation results in \emph{destructive cooperation}, where unrelated tasks negatively impact each other’s performance. To address this issue, we introduce a semantic-aware task clustering mechanism in this work that groups tasks based on their semantic similarity. This clustering is integrated within the FL process, resulting in a clustered FL approach that improves task performance.

Thus, we propose a Clustered-Federated-CMT-SemCom framework, featuring a unified encoder on each semantic transmitter (Tx) and separate task-specific decoders on the receiver (Rx) side. Semantic Txs send only locally learned parameters to a parameter server (PS), which aggregates them using clustered FL to update the shared encoders. The proposed framework is trained E2E through alternating local training and communication rounds (CRs). The main contributions are summarized as follows:

\begin{itemize}  
    \item Proposing a Clustered-Federated-CMT-SemCom framework for distributed multi-user scenarios enabling cooperative multi-tasking.
    \item Introducing a semantic-aware task clustering mechanism to mitigate destructive cooperation.
    \item Leveraging an information-theoretic criterion in the semantic domain, moving beyond common clustering approaches that rely on data distribution.
\end{itemize}

\section{System Model}\label{section:SystemModel}

We consider a multi-user wireless SemCom system with FL-enabled cooperation. The overall system consists of two logical systems, the E2E SemCom and an FL mechanism that enables cooperative distributed learning. 

The SemCom network consists of $N$ Tx-Rx pairs, each equipped with a single-antenna semantic encoder and a single-antenna semantic decoder. Each of these is associated with a downstream task. At the $i$-th Tx, the semantic encoder observes a continuous-valued source $\bm S_i$ and extracts task-specific information corresponding to the \emph{semantic variable} $Z_i \in \mathcal Z$, where all semantic variables share the same alphabet $\mathcal Z = \{c_1, c_2, \ldots, c_f\}$. The tuple $( Z_i, \bm S_i)$ forms a \emph{semantic source}, first introduced in \cite{halimi-letter} and fully described by the probability density function (pdf) $p(z_i, \bm{s}_i)$. Given an observation $\bm{s}_i$, the semantic encoder produces a channel input $\bm{x}_i$ of length $d$ according to the conditional pdf $p(\bm{x}_i|\bm{s}_i)$. The encoded signal is transmitted over an orthogonal additive white Gaussian noise (AWGN) channel. The received signal at the $i$-th decoder, after $d$ channel uses, is $\bm y_i = \bm x_i + \bm n_i$ where $\bm n_i \sim N(\bm 0_d, \gamma_n^2 \bm I_d)$. Based on $\bm{y}_i$, the semantic decoder infers an estimate $\hat{z}_i$ using $p(\hat z_i|\bm{y}_i)$ with the goal that $\hat z_i = z_i$. 
Thus, the $i$-th SemCom link induces the joint pdf
\begin{equation*}
    p(z_i,\bm{s}_i,\bm{x}_i,\bm{y}_i,\hat{z}_i) =  p(\bm{s}_i,z_i)\,p(\bm{x}_i|\bm{s}_i)\,p(\bm{y}_i|\bm{x}_i)\,p(\hat{z}_i|\bm{y}_i).
\end{equation*}

Following a data-driven approach, we consider that each SemCom Tx-Rx pair is trained from a local dataset $\mathcal D_i$, consisting of samples $(z_i,\bm{s}_i)$, available at the $i$-th SemCom Tx. Note that during transmission, only $\bm{s}_i$ is available at the encoder, while $z_i$ is used for supervised training. We adopt a variational approximation technique in which the encoder and decoder pdfs are approximated by DNNs parameterized by $\boldsymbol{\theta}_i$ and $\boldsymbol{\phi}_i$, respectively.  While all encoders share the same DNN architecture, their model parameters may differ across tasks. The decoders are task-specific, and their DNN architecture depends on the corresponding task.

Our goal is to exploit cooperative multi-tasking benefits during training without collecting the datasets $\mathcal D_i$ at a centralized location. To this end, we leverage FL where the $i$-th semantic encoder-decoder pair is trained locally from $\mathcal D_i$. This is an iterative process, where intermediate training results are synchronized using a PS at each CR. Since decoders do not share a common structure, only encoder model parameters participate in the FL process, and the decoders remain task-specific and local. Thus, a PS aggregates encoders' parameters from a subset of semantically related tasks to prevent negative transfer across heterogeneous tasks. Let $\mathcal T = \{ 1, \ldots, N \}$  denote the set of all tasks, and $\mathcal T_k \subseteq \mathcal T$ represent the $k$-th task cluster. Then, the encoder model parameters of related tasks $\{\boldsymbol{\theta}_i \mid i \in\mathcal T_k\}$ are aggregated at the PS to form a shared semantic encoder model for cluster $k$. Assuming $K$ task clusters, the PS maintains $K$ aggregated semantic encoder models. We further assume ideal communication channels between the semantic Txs and the PS during training. The aggregated encoder parameters are distributed back to the corresponding semantic encoders, enabling CMT-SemCom across related tasks.

\section{Semantic-Aware Clustered-Federated CMT-SemCom}\label{section:ClusteredF-CMT-SemCom}

We propose an efficient Clustered-Federated-CMT-SemCom for multi-user multi-task scenarios. This section introduces the Clustered-Federated-CMT-SemCom, which alternates between local training of semantic encoder-decoder pairs and periodic aggregation of $K$ unified encoders. Training is performed over multiple CRs, indexed by $t = 0, \ldots, T$. In each CR, local models are trained at the SemCom link for $E$ epochs, and the resulting parameters are transmitted to the PS for aggregation and global updating. Specifically, Section~\ref{subsec:LearningObjective} formulates an InfoMax-based objective for local training, Section~\ref{subsec:FLAggregation} describes the clustered FL-based aggregation and update procedure, and Section~\ref{subsec:Clustering} presents our semantic-aware clustering mechanism that jointly determines the number of clusters $K$ and the associated task groupings $\mathcal{T}_1, \ldots, \mathcal{T}_K$.

\subsection{Local Training Objective}\label{subsec:LearningObjective}

To design our local semantic encoder-decoder pairs, we formulate an optimization problem adopting the InfoMax principle together with the E2E learning manner as:

\begin{equation} \label{eq:infomax_opt}
        p^\star(\bm{x}_i|\bm{s}_i) =
        \argmax_{p(\bm{x}_i|\bm{s}_i)}
        I(\bm{Y}_i;Z_i),
\end{equation}
where the objective is to maximize the mutual information between the channel output $\bm{Y}_i$ and the semantic variable $Z_i$. Following similar steps as in \cite{halimi-letter}, along with the fact that the encoder pdf is approximated by DNN parameters $\boldsymbol{\theta}_i$, the approximated objective function is derived as: 

\begin{equation} \label{eq:objective_function}
    \begin{aligned}
        &\mathcal{L}_i(\boldsymbol{\theta}_i) = \, I(\bm{Y}_i; Z_i) \\[0.6em]
        &\approx  \mathbb{E}_{p(\bm{s}_i,z_i)}\left[\, \mathbb{E}_{p_{\boldsymbol{\theta}_i}(\bm{x_i|\bm{s}_i})} \left\{\mathbb{E}_{p(\bm{y}_i|\bm{x}_i)} \left[\,\log p(z_i|\bm{y}_i) \right]\ \right\} \right].
    \end{aligned}
\end{equation}

Although $p(z_i|\bm{y}_i)$ can be fully determined using the source and $p_{\boldsymbol{\theta}_i}(\bm{x}_i|\bm{s}_i)$, due to high-demensional integrals we approximate it by $p_{\boldsymbol{\phi}_i}(z_i|\bm{y}_i)$. Thus, the objective is expressed as: 

\begin{equation} \label{eq:objective_function2}
        \mathcal{L}_i(\boldsymbol{\theta}_i, \boldsymbol{\phi}_i) =  \mathbb{E}_{p(\bm{s}_i,z_i)}\left[\, \mathbb{E}_{p_{\boldsymbol{\theta}_i}(\bm{y_i|\bm{s}_i})} \left[\,\log p_{\boldsymbol{\phi}_i}(z_i|\bm{y}_i) \right]\ \right].
\end{equation}

In (\ref{eq:objective_function2}), we emphasize the joint semantic and channel coding performed by the encoders as we consider $p_{\boldsymbol{\theta}_i}(\bm{y}_i|\bm{s}_i)=\int p_{\boldsymbol{\theta}_i}(\bm{x}_i|\bm{s}_i)\,p(\bm{y}_i|\bm{x}_i)\,d\bm{x}_i$. It also reflects our adaptation of E2E design, where we jointly optimize the encoders and decoders, including the wireless channel, as its transfer function is differentiable.

By replacing the source distribution \mbox{$p(\bm{s}_i,z_i)$} with the corresponding training data set \mbox{$\mathcal D_i = \{ (\bm s_i^{(b)}, z_i^{(b)}) \}_{b = 1}^{|\mathcal D_i|}$} of size $|\mathcal D_i|$, and approximating the expectations using Monte Carlo sampling, we obtain the empirical estimate of the above objective function
\begin{equation} \label{eq:empirical_objective}
        \mathcal{L}_i(\boldsymbol{\theta_i}, \boldsymbol{\phi_i}) \approx \frac{1}{|\mathcal{D}_i|}\sum_{b=1}^{|\mathcal{D}_i|}\left[\,\frac{1}{L}\sum_{l=1}^{L}\bigg[\, \log p_{\boldsymbol{\phi}_i}(z_b|\bm{y}_{b,l})\bigg]\ \right],
\end{equation}
where we fix the sample size of the encoder output and the channel sampling $\bm{y}_{b,l} = \bm{x}_{b} + \bm{n}_l$, equal for each batch.

\subsection{Clustered FL Aggregation}\label{subsec:FLAggregation}

With the local learning objectives in place, we now present the clustered FL aggregation and updating scheme for cooperative multi-tasking in multi-user environments. During each CR, two steps are performed iteratively. In the first step, the semantic encoder-decoder pairs are locally optimized based on the objective in~(\ref{eq:empirical_objective}). In the second step, at the $t$-th CR, all locally trained encoders' parameters $\{\boldsymbol{\theta}^{(t-1)}\}_i$ are transmitted to the PS, and the aggregation takes place to update the $K$ unified semantic encoders as follows:

\begin{equation} \label{eq:FL_aggregation} 
\bar{\boldsymbol{\theta}}_{k}
=
\frac{1}{|\mathcal{T}_k|}
\sum_{i\in \mathcal{T}_k}
\,\boldsymbol{\theta}^{\star}_i
\qquad k = 1,\ldots,K. 
\end{equation}

In~(\ref{eq:FL_aggregation}), $\bar{\boldsymbol{\theta}}_{k}$ represents the aggregated semantic encoder parameters shared by all semantic encoders within the $k$-th task cluster. Clusters $\{\mathcal T_k\}_{k=1}^{K^\star}$, are constructed so that $\bigcup_{k=1}^{K} \mathcal{T}_k = \mathcal{T}$, ensuring that all participating encoders are included in the aggregation process, while $\mathcal{T}_k \cap \mathcal{T}_{k'} = \emptyset$ for all $k \neq k'$, which enforces a disjoint cluster structure. 

Consequently, the aggregation in \eqref{eq:FL_aggregation} corresponds to a cluster-wise averaging of the locally trained encoder parameters, enabling \emph{constructive and cooperative multi-tasking}. In Section~\ref{subsec:Clustering}, we present the semantic-aware task clustering approach to determine optimal $K$ and $\mathcal{T}_1, \ldots,\mathcal{T}_K$.

\subsection{Semantic-Aware Task Clustering}\label{subsec:Clustering}

Our goal is to characterize the relationships among heterogeneous tasks such that personalized global models are learned, as shown in (\ref{eq:FL_aggregation}), in order to guarantee constructive and cooperative multi-tasking among groups of semantic links. Therefore, for efficient MTL, we aim to group tasks based on their \emph{semantic relatedness}, where semantic relatedness is defined by the similarity of the distributions of the underlying semantic variables $\bm{Z}=[\,Z_1, \ldots, Z_N ]\,$. 

A common information-theoretic measure of distributional similarity is the Kullback-Leiber (KL) divergence $D(\mathds P\| \mathds Q)$ with $\mathds P$ and $\mathds Q$ being arbitrary distributions. This suggests a clustering approach that groups tasks with mutually large KL-divergence together. However, a clustering metric should be symmetric and bounded. One symmetrization of the KL-divergence with these properties is the Jensen-Shannon (JS) divergence \cite{Lin1991}, defined as
\begin{equation}
    D_{JS} (\mathds P\|\mathds Q) = \frac{1}{2} D(\mathds P\|\mathds M) + 
    \frac{1}{2} D(\mathds Q\|\mathds M),
\end{equation}
where  $\mathds M$ is the mixture distribution $\frac{\mathds P+ \mathds Q}{2}$. The JS-divergence is nonnegative, takes the value zero if and only if $\mathds P = \mathds Q$, and is upper-bounded by $\log 2$.
Since the JS-divergence quantifies how similar two distributions are to their average, it implies that these distributions become more similar to each other.

Computing the JS-divergence requires knowledge of the true distribution of $\bm Z$. Since this information is unavailable, we obtain the empirical probability mass function (pmf) from $\mathcal D$ instead. Let the semantic variables alphabet, $\mathcal Z$, then, for $|\mathcal{D}_i|$ large enough,
\begin{equation} \label{eq:PMF}
	p_{Z_i}(c) \approx \pi_i(c) = \frac{\sum_{\bm z \in \mathcal D_i} \bm 1(z_i = c)}{|\mathcal{D}_i|},
\end{equation}
for all $i$ and $c \in\mathcal Z$. These empirical pmfs are computed on-device and are low-dimensional in comparison to the data dimension. The semantic Txs transmit their empirical $\pi_{Z_i}$ to the PS prior to the FL training process. 

Next, we define the empirical semantic similarity $\omega_{ij}$ between $Z_i$ and $Z_j$, computed on the PS, as
\begin{equation}\label{eq:similarity}
	\omega_{ij} = 1 - \frac{1}{2} \sum_{c \in \mathcal{Z}} \left( \pi_i(c) \log\frac{\pi_i(c)}{A_{ij}(c)} + \pi_j(c) \log\frac{\pi_j(c)}{A_{ij}(c)} \right),
\end{equation}
with $A_{ij}(c)=\frac{1}{2} (\pi_i(c)+\pi_j(c))$. Due to the law of large numbers, $\omega_{ij} \to 1 - D_{JS}(Z_i \| Z_j)$ as $|\mathcal{D}|\to\infty$. Clearly, $\omega_{ij} = \omega_{ji}$ and $0 \le \omega_{ij} \le 1$ with $\omega_{ij} = 1$ if and only if $Z_i$ and $Z_j$ have the same distribution. Thus, tasks with semantic similarity $\omega_{ij}$ close to 1 should be grouped together, while tasks with $\omega_{ij}$ close $0$ must not. We define $\boldsymbol{\Omega} \in [0,1]^{N \times N}$ as the semantic similarity matrix with entries $\omega_{ij}$. 

The clustering can be interpreted as a fully connected, undirected graph $\mathcal G$ with $N$ vertices representing $Z_1, Z_2, \ldots, Z_N$ and edge weights $\omega_{ij}$. We want to cluster our related semantic variables such that each cluster keeps its full connectivity and no overlaps occur between the clusters. In other words, we want to remove all edges from $\mathcal G$ with small $\omega_{ij}$ and then identify the minimum number of fully connected subgraphs.

To induce this, we introduce a similarity threshold $\tau \in (0,1)$, which acts as a design hyperparameter. Based on this $\tau$, we define a binary task–task relation matrix $\mathbf R \in \{0,1\}^{N \times N}$ based on $\boldsymbol{\Omega}$, as its elements $r_{ij} = 1$ if $\omega_{ij} \ge \tau$, and $0$ otherwise. To avoid overlaps, if task $i$ is grouped with task $j$, and task $j$ is grouped with task $k$, then task $i$ must also be grouped with task $k$. Otherwise, tasks $i$ and $k$ could belong to different clusters that both include task $j$, leading to overlap. To enforce this, we impose the transitivity constraint: $r_{ij} + r_{jk} \leq r_{ik} + 1$. This results in a block matrix of $\mathbf{R}$, where the optimal number of clusters, $K^\star$, is identified by extracting the connected components in this matrix. 

Further, the optimum clusters $\{\mathcal T_k\}_{k=1}^{K^\star}$ are also determined, without the need to transmit the local datasets to the PS. As a result, the clustered FL aggregation shown in (\ref{eq:FL_aggregation}) is implemented based on the semantic similarities, without requiring a priori specification of $K$. The overall scheme of the proposed framework is presented in Algorithm \ref{algorithm:Overall}.

\section{Simulation Results} \label{section:Simulation}

\algrenewcommand\algorithmicrequire{\textbf{Input:}}
\algrenewcommand\algorithmicensure{\textbf{Output:}}
\begin{algorithm}[!t]
    \footnotesize
    \caption{The Overall Training Process of the Clustered-Federated-CMT-SemCom.} \label{algorithm:Overall}
    \begin{algorithmic}[1]
        \Require Local datasets $\{\mathcal{D}_i\}_{i=1}^N$, communication rounds $T$, local epochs $E$
        \Ensure Number of clusters $K^\star$, clusters $\{\mathcal T_k\}_{k=1}^{K^\star}$, shared encoders $\{\bar{\boldsymbol{\theta}}^\star_k\}_{k=1}^{K^\star}$, task-specific decoders $\{\boldsymbol{\phi}^\star_i\}_{i=1}^{N}$
        
        \Statex \textbf{Phase 1:} Semantic-Aware Task Clustering
        \For{SemCom link $i = 1$ to $N$}
            \State Compute empirical semantic pmf $\pi_{Z_i}$ based on (\ref{eq:PMF}).
            \State Transmit low-dimensional $\pi_{Z_i}$ to the PS.
        \EndFor
        \State PS calculates semantic similarities by (\ref{eq:similarity}) using $\{\pi_{Z_i}\}_{i=1}^N$.
        \State Given the similarity threshold $\tau$, task-task relation matrix $\mathbf{R}$ is calculated.
        \State The clusters $\{\mathcal T_k\}_{k=1}^{K^\star}$ are identified at the PS by the block matrix $\mathbf{R}$.
        \vspace{0.3em}
        \Statex \textbf{Phase 2:} Proposed Federated Training Process
        \While{communication rounds $t=0$ to $T$}
        \For{SemCom link $i=1$ to $N$}
            \For{$e=1$ to $E$}
            \State $\bm{x}_i \leftarrow p_{\boldsymbol{\theta}_i}(\bm{x_i}|\bm{s}_i)$.
            \State Transmit $\bm{x}_i$ over wireless channel and get $\bm{y}_i$.
            \State $\hat{z}_i \leftarrow p_{\boldsymbol{\phi}_i}(z_i|\bm{y_i})$.
            \State Calculate the loss $\mathcal{L}_i(\boldsymbol{\theta}_i, \boldsymbol{\phi}_i)$ based on (\ref{eq:empirical_objective}).
            \EndFor
        \State Transmit the locally trained encoder $\boldsymbol{\theta}^\star_i$ to the PS.
        \EndFor
        \State PS aggregates models based on (\ref{eq:FL_aggregation}).
        \State PS distributes $\{\bar{\boldsymbol{\theta}}^\star_k\}_{k=1}^{K^\star}$ to the corresponding semantic Txs.
        \EndWhile
    \end{algorithmic}
\end{algorithm}
 
\begin{figure}[!t]
    \centering
    \resizebox{0.4\textwidth}{!}{\input{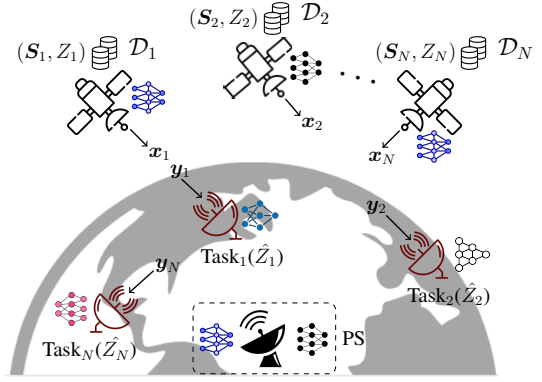}}
    \caption{Illustration of the proposed Clustered-Federated-CMT-SemCom framework applied to LEO satellite network.}
    \label{fig:sys_model}
\end{figure}

To demonstrate the effectiveness of the proposed approach, we apply it to a LEO satellite network scenario, as a representative use case, Fig.~\ref{fig:sys_model}. We utilize the MNIST dataset of handwritten digits, which contains 60,000 images for the training set and 10,000 images for the test set. The dataset is pre-processed by assigning multiple labels to each data sample $\bm s_i$, as we focus on a multi-label domain MTL. Thus, the label domain defines our $\mathcal Z$ and each label corresponds to a semantic variable $Z_i$. Accordingly, the local datasets $\mathcal D_i$ available at each satellite, referred to as ``Sat" in this section, is constructed as $\mathcal D_i = \{(\bm{s}^{(b)}_i, z^{(b)}_i)\}_{b=1}^{|D_i|}$.

We consider three learning tasks: two binary classification tasks, classification of digit ``$2$" as ``Task1", and classification of digit ``$6$" as ``Task3" in our setup. And one categorical classification task, which is the digit identification of $\{1, 2\}$ as ``Task2". Therefore, the semantic variables are $\{z_1, z_3\}$ $\sim$ Bernoulli and $z_2$ $\sim$ Multinomial distributions. 

To emulate realistic deployment conditions, the training data are distributed heterogeneously across the Sats, while maintaining approximately equal local dataset sizes, i.e., $\mathcal |D_1| \simeq \mathcal |D_2| \simeq \mathcal |D_3|$. All Sats use identical two-layer fully connected semantic encoders with ReLU activations and an output dimension of $8$. Decoders are task-specific and differ only in their output dimensions, while the PS maintains the encoders architectures for aggregation without a decoder. 

The number of DNN encoders instantiated at the PS depends on the number of task clusters $K^\star$. The task similarity threshold $\tau$, is determined by considering the middle points between the lowest and highest similarity scores in $\boldsymbol{\Omega}$. Then, based on the proposed semantic-aware clustering method described in Section~\ref{subsec:Clustering}, we obtain $K^\star = 2$, where Task1 and Task2 form a shared cluster, and Task3 constitutes a separate cluster. The training procedure consists of $30$ CRs, each comprising $10$ local training epochs using the loss function in (\ref{eq:empirical_objective}), resulting in a total of $300$ local training epochs per Sat.

To assess the performance of our framework, named as Clustered FL-CMT-SemCom, we compare its task execution accuracy against two baseline schemes: Unclustered FL-CMT-SemCom and Individual training. In the Unclustered FL-CMT-SemCom, locally trained model parameters are aggregated in a task-agnostic manner, without accounting for task clustering. The Individual training corresponds to single-task learning, where each task is trained independently, and no MTL is performed.

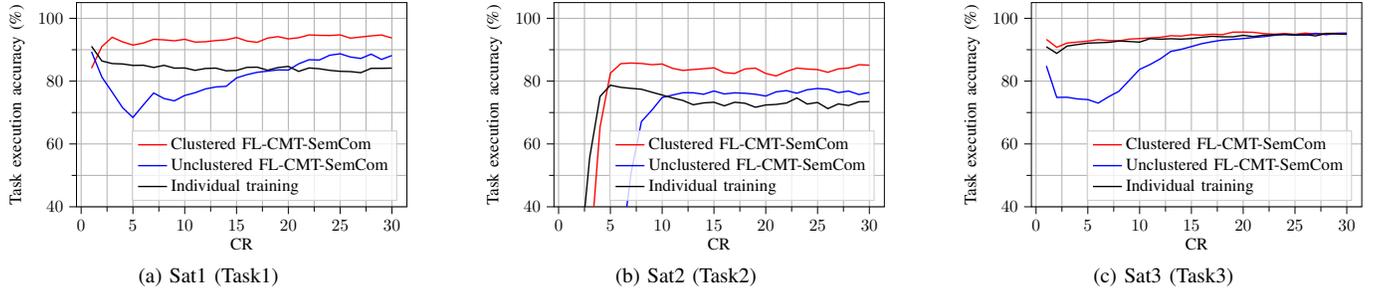
\begin{figure*}[t!]
    \centering
    \subfloat[Sat1 (Task1)]{%
        \resizebox{0.3\textwidth}{!}{
\begin{tikzpicture}

\definecolor{darkgrey176}{RGB}{176,176,176}
\definecolor{darkorange25512714}{RGB}{255,127,14}
\definecolor{forestgreen4416044}{RGB}{44,160,44}
\definecolor{lightgrey204}{RGB}{204,204,204}
\definecolor{steelblue31119180}{RGB}{31,119,180}

\begin{axis}[
legend cell align={left},
legend style={
  fill opacity=0.8,
  draw opacity=1,
  text opacity=1,
  at={(0.97,0.03)},
  anchor=south east,
  draw=lightgrey204
},
tick align=outside,
tick pos=left,
grid=both,
grid style={line width=.2pt, draw=gray!10},
minor grid style = {line width=.2pt},
minor x tick num = 1,
minor y tick num = 1,
x grid style={darkgrey176},
xlabel={CR},
xmin=-0.45, xmax=31.45,
xtick style={color=black},
width = \axisdefaultwidth,
height = 0.8*\axisdefaultheight,
y grid style={darkgrey176},
ylabel={Task execution accuracy (\%)},
ymin=40, ymax=105,
ytick style={color=black}
]
\addplot [thick, red]
table {%
1 84.13
2 91.01
3 93.96
4 92.5
5 91.48
6 92.11
7 93.32
8 93.14
9 92.84
10 93.32
11 92.42
12 92.54
13 92.92
14 93.15
15 93.89
16 92.77
17 92.36
18 93.7
19 94.17
20 93.45
21 93.82
22 94.71
23 94.58
24 94.52
25 94.73
26 93.68
27 94.05
28 94.41
29 94.69
30 93.72
};

\addlegendentry{Clustered FL-CMT-SemCom}

\addplot [thick, blue]
table {%
1 89.37
2 81.29
3 76.47
4 71.55
5 68.43
6 72.41
7 76.23
8 74.46
9 73.73
10 75.44
11 76.31
12 77.52
13 78.15
14 78.35
15 81.02
16 82.05
17 82.81
18 83.19
19 83.58
20 83.52
21 85.51
22 86.81
23 86.72
24 88.25
25 88.71
26 87.67
27 87.22
28 88.59
29 86.91
30 88.15
};
\addlegendentry{Unclustered FL-CMT-SemCom}

\addplot [thick, black]
table {%
1 91.1
2 86.42
3 85.59
4 85.47
5 85
6 85.06
7 84.35
8 85.03
9 84.14
10 84.19
11 83.44
12 84.03
13 84.15
14 83.29
15 83.41
16 84.38
17 84.44
18 83.48
19 84.32
20 84.69
21 83.1
22 84.19
23 83.94
24 83.43
25 83.13
26 83.04
27 82.69
28 84.07
29 84.08
30 84.14
};
\addlegendentry{Individual training}
\end{axis}

\end{tikzpicture}}%
        \label{fig:sat1}%
    }%
    \hfill%
    \subfloat[Sat2 (Task2)]{%
        \resizebox{0.3\textwidth}{!}{
\begin{tikzpicture}

\definecolor{darkgrey176}{RGB}{176,176,176}
\definecolor{darkorange25512714}{RGB}{255,127,14}
\definecolor{forestgreen4416044}{RGB}{44,160,44}
\definecolor{lightgrey204}{RGB}{204,204,204}
\definecolor{steelblue31119180}{RGB}{31,119,180}

\begin{axis}[
legend cell align={left},
legend style={
  fill opacity=0.8,
  draw opacity=1,
  text opacity=1,
  at={(0.97,0.03)},
  anchor=south east,
  draw=lightgrey204
},
tick align=outside,
tick pos=left,
grid=both,
grid style={line width=.2pt, draw=gray!10},
minor grid style = {line width=.2pt},
minor x tick num = 1,
minor y tick num = 1,
x grid style={darkgrey176},
xlabel={CR},
xmin=-0.45, xmax=31.45,
xtick style={color=black},
width = \axisdefaultwidth,
height = 0.8*\axisdefaultheight,
y grid style={darkgrey176},
ylabel={Task execution accuracy (\%)},
ymin=40, ymax=105,
ytick style={color=black}
]
\addplot [thick, red]
table {%
1 17.75
2 20.52
3 22.8
4 65.39
5 82.53
6 85.56
7 85.76
8 85.62
9 85.23
10 85.45
11 84.07
12 83.42
13 83.7
14 83.95
15 84.23
16 82.71
17 82.45
18 83.89
19 84.13
20 82.44
21 81.65
22 83.06
23 84.16
24 83.85
25 83.67
26 82.81
27 83.9
28 84.18
29 85.23
30 85.07
};

\addlegendentry{Clustered FL-CMT-SemCom}

\addplot [thick, blue]
table {%
1 19.29
2 18.4
3 19.46
4 20.24
5 20.73
6 24.9
7 50.83
8 67.15
9 70.74
10 74.8
11 75.57
12 76.31
13 76.29
14 75.82
15 76.86
16 75.92
17 76.28
18 76.14
19 75.82
20 75.25
21 76.59
22 76.98
23 76.15
24 77.26
25 77.65
26 77.42
27 76.35
28 76.83
29 75.74
30 76.41
};
\addlegendentry{Unclustered FL-CMT-SemCom}

\addplot [thick, black]
table {%
1 18.96
2 21.04
3 55.66
4 75.11
5 78.71
6 78.03
7 77.69
8 77.42
9 76.48
10 75.58
11 74.65
12 73.85
13 72.48
14 73.07
15 73.28
16 72.14
17 73.28
18 72.96
19 71.69
20 72.41
21 72.58
22 73.03
23 74.65
24 72.71
25 73.21
26 71.25
27 72.75
28 72.24
29 73.41
30 73.5
};
\addlegendentry{Individual training}
\end{axis}

\end{tikzpicture}}%
        \label{fig:sat2}%
    }%
    \hfill%
    \subfloat[Sat3 (Task3)]{%
        \resizebox{0.3\textwidth}{!}{
\begin{tikzpicture}

\definecolor{darkgrey176}{RGB}{176,176,176}
\definecolor{darkorange25512714}{RGB}{255,127,14}
\definecolor{forestgreen4416044}{RGB}{44,160,44}
\definecolor{lightgrey204}{RGB}{204,204,204}
\definecolor{steelblue31119180}{RGB}{31,119,180}

\begin{axis}[
legend cell align={left},
legend style={
  fill opacity=0.8,
  draw opacity=1,
  text opacity=1,
  at={(0.97,0.03)},
  anchor=south east,
  draw=lightgrey204
},
tick align=outside,
tick pos=left,
grid=both,
grid style={line width=.2pt, draw=gray!10},
minor grid style = {line width=.2pt},
minor x tick num = 1,
minor y tick num = 1,
x grid style={darkgrey176},
xlabel={CR},
xmin=-0.45, xmax=31.45,
xtick style={color=black},
width = \axisdefaultwidth,
height = 0.8*\axisdefaultheight,
y grid style={darkgrey176},
ylabel={Task execution accuracy (\%)},
ymin=40, ymax=105,
ytick style={color=black}
]
\addplot [thick, red]
table {%
1 93.29
2 90.77
3 92.13
4 92.43
5 92.76
6 93.19
7 92.94
8 92.84
9 93.38
10 93.54
11 93.74
12 93.96
13 94.45
14 94.34
15 94.79
16 94.63
17 94.94
18 94.79
19 95.59
20 95.62
21 95.54
22 95.13
23 94.94
24 95.16
25 94.9
26 95.3
27 95.01
28 95.21
29 95.25
30 95.28
};

\addlegendentry{Clustered FL-CMT-SemCom}

\addplot [thick, blue]
table {%
1 84.84
2 74.82
3 74.87
4 74.34
5 74.13
6 73.01
7 75.02
8 76.76
9 80.25
10 83.74
11 85.27
12 87.1
13 89.44
14 90.11
15 91.01
16 91.93
17 92.55
18 93.07
19 93.31
20 93.55
21 93.89
22 94.18
23 94.58
24 94.89
25 94.66
26 94.68
27 95.22
28 94.79
29 95.19
30 95.22
};
\addlegendentry{Unclustered FL-CMT-SemCom}

\addplot [thick, black]
table {%
1 90.92
2 88.84
3 91.16
4 91.65
5 92.11
6 92.22
7 92.35
8 92.75
9 92.62
10 92.42
11 93.52
12 93.35
13 93.5
14 93.37
15 93.55
16 93.97
17 94.28
18 94.11
19 94.1
20 94.58
21 94.2
22 94.63
23 94.77
24 94.83
25 94.69
26 94.81
27 94.41
28 95.15
29 95
30 94.9
};
\addlegendentry{Individual training}
\end{axis}

\end{tikzpicture}}%
        \label{fig:sat3}%
    }%
    \caption{Performance comparison of the proposed framework vs. the benchmarks: Unclustered FL-CMT-SemCom and Individual training on the test set.}
    \label{fig:Performace}
\end{figure*}

\begin{figure}[!t]
    \centering
    \resizebox{0.4\textwidth}{!}{\pgfmathsetmacro{\avgA}{(93.72+85.07+95.28)/3}
\pgfmathsetmacro{\avgB}{(88.15+76.41+95.22)/3}
\pgfmathsetmacro{\avgC}{(84.14+73.5+94.9)/3}

\begin{tikzpicture}
\begin{axis}[
    ybar,
    bar width=20pt,
    width = \axisdefaultwidth,
    height = 0.75*\axisdefaultheight,
    ymin=80, ymax=100,
    ylabel={Task execution accuracy (\%)},
    xtick=\empty, 
    enlarge x limits=0.8,
    nodes near coords={\pgfmathprintnumber[fixed,precision=2]{\pgfplotspointmeta}},
    grid=both,
    legend style={
        at={(0.98,0.98)}, 
        anchor=north east,
        draw=black, 
        fill=white, 
        font=\small,
    }
]

\addplot[fill=red, draw=red] coordinates {(1, \avgA)};
\addlegendentry{Clustered FL-CMT-SemCom}

\addplot[fill=blue, draw=blue] coordinates {(2, \avgB)};
\addlegendentry{Unclustered FL-CMT-SemCom}

\addplot[fill=black, draw=black] coordinates {(3, \avgC)};
\addlegendentry{Individual training}

\end{axis}
\end{tikzpicture}}
    \caption{Comparison of the overall network performance of the approaches.}
    \label{fig:Overall_performance}
\end{figure}
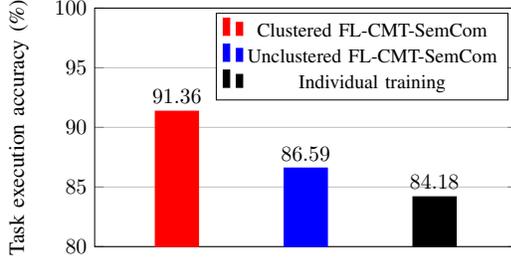

Fig.~\ref{fig:sat1} and~\ref{fig:sat2} show the rapid improvement and peak accuracy achieved by the proposed framework. These results demonstrate that cooperative multi-tasking improves task execution performance compared to individual training when semantically related tasks are jointly processed. Fig.~\ref{fig:sat3} highlights the destructive cooperation in the unclustered setting where task-agnostic parameter aggregation leads to performance degradation relative to individual setting. We also observe that the proposed method is performing the same as the individual one, as it has been clustered alone. Finally, Fig.~\ref{fig:Overall_performance} presents the average performance across all tasks, showing that the proposed Clustered FL-CMT-SemCom consistently improves overall task execution accuracy across the network. 

In addition, the results demonstrate that the proposed approach requires less FL communication exchange between the Sats and the PS to achieve higher accuracy, highlighting the communication efficiency it brings to the system.

\section{Conclusion} \label{section:Conclusion}
We extended CMT-SemCom to distributed multi-user scenarios, such as in non-terrestrial satellite networks, by introducing Clustered-Federated-CMT-SemCom. FL enables cooperative multi-tasking across distributed users, while an information-theoretic, semantic-aware task clustering ensures constructive cooperation among heterogeneous tasks. Simulation results show accuracy gains of up to $7.18\%$ and $4.77\%$ over the Unclustered FL-CMT-SemCom and Individual training benchmarks, respectively, along with faster convergence. Future work includes incorporating realistic wireless channels, studying the impact of noisy model updates, and developing more systematic semantic-aware clustering methods.

\balance
\bibliographystyle{IEEEtran}
\bibliography{IEEEabrv,References.bib}

\end{document}